\documentclass[reqno]{amsbook}%
\usepackage{setspace}
\usepackage{amsmath}%
\usepackage{amsfonts}%
\usepackage{amssymb}%
\usepackage{graphicx}
\usepackage{microtype}

\theoremstyle{plain}

\numberwithin{equation}{section}
\begin{document}
\onehalfspacing
\raggedbottom
\frontmatter
\title[Beta-Like Distributions]{Exploring Beta-Like Distributions}
\author[H.R.N. van Erp]{H.R.N. van Erp}
\author{R.O. Linger}
\author{P.H.A.J.M. van Gelder}

\begin{abstract}
The most well known probability distribution of probabilities is the Beta distribution. If we have observed $r$ `successes', each having a probability $\theta$, and $n-r$ `failures', each having a probability $1-\theta$. In this paper we will derive a whole family of Beta-like distributions, which take as their data not only the number of successes and failures, but also values on predictor variables and time to failure or time without failure. \end{abstract}
\maketitle
\tableofcontents

\chapter*{Preface}
The most well known probability distribution of probabilities is the Beta distribution. If we have observed $r$ `successes', each having a probability $\theta$, and $n-r$ `failures', each having a probability $1-\theta$. Then the corresponding Beta distribution if $\theta$ is given as:
\[
	p\!\left(\left.\theta\right|r,n\right) = \frac{\left(n-1\right)!}{\left(r-1\right)! \left(n-r-1\right)!} \theta^{r-1} \left(1-\theta\right)^{n-r-1}.
\]

We will proceed in this paper to derive a whole family of Beta-like distributions, which take as their data not only the number of successes and failures, but also values on predictor variables and time to failure or time without failure. 

The recurring theme in all this will be that, apart from the ordinary product and sum rules (\textit{e.g.}, Bayes' theorem and the integrating out of nuisance parameters), a change-of-variable or, alternatively, a Jacobian transformation allows us to map the uncertainty we have, regarding the unknown parameter(s), as captured in the corresponding posterior, unto the probability of interest; thus, allowing us to construct a probability distribution of the probability of interest.

The Beta-Like distribution is the distribution that takes into account the epistemological parameter uncertainty, as captured in the posterior distribution of these parameters, of the parameters of a given probability model. The Bayesian model selection, also discussed in this paper, takes into account the epistemological model uncertainty\footnote{In the Bayesian view of probability theory there is no uncertainty other than epistemological, seeing that a probability distribution over some set of propositions reflect our state of knowledge regarding the plausibilities of these propositions, \cite{Jaynes03}. For example, the coin has a mass, a center of gravity, a circumference, a width, etc... . But it does not have the physical property: the probability of head or tails. And we know of at least one recorded instance were a coin landed spinning on its side and remained standing on its side as its spinning subsided, until it came to a halt, while still standing on its side.}.

In this paper we will be talking about failure mechanisms which have some underlying probabilistic `generating' process. In light of this loose terminolgy, we would like to give, as a caveat, the following quote\footnote{In this quote Jaynes answers the charge that only long term frequencies of random experiments may be considered `objective'. Bayesian probability theory was formulated in 1774 by the physicist Laplace, who used this probability theory to identify those problems in celestial mechanics where the data seemed to contradict the then current theory. This allowed Laplace to be highly productive in this field of science, so much so, that in his time he was called the French Newton. But shortly after his death Laplace's probability theory was attacked by a school of pure mathematicians, who thought the definition of a probability as a state of knowledge to be lacking in rigor. Instead, they proposed that a probability should mean to be the `observed' long term frequency of (an imaginary infinity of) random experiments. For a time this viewpoint dominated the field so completely that those who were students in the period 1930-1960 were hardly aware that any other conception had ever existed, \cite{Jaynes76}.} by Jaynes, who is considered by many to be the father of modern Bayesianity, \cite{Jaynes76}:
\begin{quotation}
	[T]he judgment of a competent engineer, based on data of past experience in the field, represents information fully as `objective' and reliable as anything we can possibly learn from a random experiment. Indeed, most engineers would make a stronger statement; since a random experiment is, by definition, one in which the outcome - and therefore the conclusion we draw from it - is subject to uncontrollable variations, it follows that the only fully `objective' means of judging the reliability of a system is through analysis of stresses, rate of wear, etc., which avoids random experiments altogether.
	
In practice, the real function of a reliability test is to check against the possibility of completely unexpected modes of failure; once a given failure mode is recognized and its mechanism understood, no sane engineer would dream of judging its chances of occurring merely from a random experiment.	   
\end{quotation}

In closing, the probability models used in this paper are by no way exhaustive. For example, we are currently studying the Negative Binomial probability model. A probability model which is particularly popular among seismologists. And we have already found that a lot of interesting things can be said about this about the Negative Binomial generating process. But this will be subject of another paper.

\mainmatter


\chapter{Explicit Probability Distributions for Probabilities}

\section{Predictors, the Logistic Regression Model.}

\subsection{The Probability Model.}
In the logistic regression model, we model the log-odds of some event by way of a regression model, say,
\begin{equation}
	\label{eq.1.2}
	\log \frac{\theta}{1-\theta} = \beta_{0} + \beta_{1} z,
\end{equation}
where $z$ is some predictor value. Identity \eqref{eq.1.2} implies that the probability of success $\theta$ can be written down as as the following function of the unknown parameters $\beta_{0}$ and $\beta_{1}$:
\begin{equation}
	\label{eq.1.3}
	\theta = \frac{\exp\left(\beta_{0} + \beta_{1} z\right)}{1 + \exp\left(\beta_{0} + \beta_{1} z\right)}.
\end{equation}
Likewise, the probability of a failure can be written down as:
\begin{equation}
	\label{eq.1.4}
	1 - \theta = \frac{1}{1 + \exp\left(\beta_{0} + \beta_{1} z\right)}.
\end{equation}

\subsection{The Likelihood, Prior, and Posterior.}
Say, we have $r$ successes, with corresponding predictor values $x_{1}, \ldots, x_{r}$, and $n-r$ failures, with corresponding predictor values $y_{1}, \ldots, y_{n-r}$. Then, by way of \eqref{eq.1.3} and \eqref{eq.1.4}, the probability of the observed data, or, equivalently, the likelihood of the unknown parameters $\beta_{0}$ and $\beta_{1}$, can be written down as:
\begin{equation}
	\label{eq.1.5}
	p\!\left(\left.D\right|\beta_{0}, \beta_{1}\right) = L\!\left(\beta_{0}, \beta_{1}\right) = \prod_{i=1}^{r} \frac{\exp\left(\beta_{0} + \beta_{1} x_{i}\right)}{1 + \exp\left(\beta_{0} + \beta_{1} x_{i}\right)} \prod_{j=1}^{n-r} \frac{1}{1 + \exp\left(\beta_{0} + \beta_{1} y_{j}\right)}
\end{equation}

We assign some uniform prior to the unknown parameters $\beta_{0}$ and $\beta_{1}$, as is customary in Bayesian regression analysis, \cite{Zellner72}:
\begin{equation}
	\label{eq.1.6}
	p\!\left(\beta_{0}, \beta_{1}\right) \propto \text{constant}, 
\end{equation}  
where `$\propto$' is the proportionality sign. 

By multiplying the likelihood with the prior, respectively, \eqref{eq.1.5} and \eqref{eq.1.6}, one may obtain, by way of product rule, or, equivalently, Bayes' theorem, \cite{Jaynes03}, the joint distribution of $\beta_{0}$ and $\beta_{1}$:
\begin{equation}
	\label{eq.1.7}
	p\!\left(D, \beta_{0}, \beta_{1}\right) \propto \prod_{i=1}^{r} \frac{\exp\left(\beta_{0} + \beta_{1} x_{i}\right)}{1 + \exp\left(\beta_{0} + \beta_{1} x_{i}\right)} \prod_{j=1}^{n-r} \frac{1}{1 + \exp\left(\beta_{0} + \beta_{1} y_{j}\right)}.
\end{equation}

Now, if we wish to obtain the probability distribution of the unknown parameters $\beta_{0}$ and $\beta_{1}$, conditional on the data $D$, or, equivalently, the posterior of $\beta_{0}$ and $\beta_{1}$, we must, by way of the product rule, \cite{Jaynes03}, divide \eqref{eq.1.7} with the evidence
\begin{equation}
	\label{eq.1.8}
	p\!\left(D\right) = \int \int p\!\left(D, \beta_{0}, \beta_{1}\right) d\beta_{0} d\beta_{1} = C.
\end{equation}
However, if we do this, then the inverse of the evidence, $C^{-1}$, being a constant not dependent upon the parameters $\beta_{0}$ and $\beta_{1}$, gets absorbed in the proportionality sign of \eqref{eq.1.7}; thus, giving us a posterior:
\begin{equation}
	\label{eq.1.9}
	p\!\left(\left.\beta_{0}, \beta_{1}\right|D\right) = \frac{p\!\left(D, \beta_{0}, \beta_{1}\right)}{p\!\left(D\right)} \propto \prod_{i=1}^{r} \frac{\exp\left(\beta_{0} + \beta_{1} x_{i}\right)}{1 + \exp\left(\beta_{0} + \beta_{1} x_{i}\right)} \prod_{j=1}^{n-r} \frac{1}{1 + \exp\left(\beta_{0} + \beta_{1} y_{j}\right)},
\end{equation}
which is proportional to both the likelihood \eqref{eq.1.5} and the joint distribution \eqref{eq.1.7}.

However, we are not that much interested in the probability distribution of $\beta_{0}$ and $\beta_{1}$. Since we are aiming for the probability distribution of the probability of a success $\theta$, given a predictor value $z$, \eqref{eq.1.3}.

Would we know the values of $\beta_{0}$ and $\beta_{1}$ exactly, as we know our predictor value $z$, then we could substitute these values directly into \eqref{eq.1.3} and, thus, get the exact probability $\theta$. Now, we do not know the values of $\beta_{0}$ and $\beta_{1}$ exactly. Instead, we have a range of probable values on the $\beta_{0}$- and $\beta_{1}$-axes, as captured by the posterior \eqref{eq.1.9}. This corresponds, through a two-to-one mapping, with a range of probable values on the $\theta$-axis. This two-to-one mapping is, typically, accomplished by way of a Jacobian transformation.

\subsection{The Jacobian Transformation.}
By way of \eqref{eq.1.3}, we have that
\[
	\theta = \frac{\exp\left(\beta_{0} + \beta_{1} z\right)}{1 + \exp\left(\beta_{0} + \beta_{1} z\right)}.
\]
So, a possible transformation from $\left(\beta_{0}, \beta_{1}\right)$ to $\left(\theta, b_{1}\right)$ is
\begin{equation}
	\label{eq.1.10}
	\beta_{0} = -\log\frac{1-\theta}{\theta} - b_{1} z, \qquad \beta_{1} = b_{1},
\end{equation}
which has a corresponding Jacobian of
\begin{equation}
	\label{eq.1.11}
	J = 
\left|\begin{array}{cc}
	\frac{\partial}{\partial \theta} \beta_{0} & \frac{\partial}{\partial b_{1}} \beta_{0} \\
	\frac{\partial}{\partial \theta} \beta_{1} & \frac{\partial}{\partial b_{1}} \beta_{1} 
\end{array}\right| = \left|\begin{array}{cc}
	\frac{1}{\theta\left(1-\theta\right)} & -z \\
	0 & 1
\end{array}\right| = \frac{1}{\theta\left(1-\theta\right)}.
\end{equation}

Substituting \eqref{eq.1.10} into the posterior \eqref{eq.1.9}, and multiplying it with the Jacobian \eqref{eq.1.11}, gives us the transformed posterior:
\begin{equation}
	\label{eq.1.12}
	p\!\left(\left.\theta, b_{1}\right|z, D\right) \propto \frac{1}{\theta\left(1-\theta\right)} \prod_{i=1}^{r} \frac{\frac{\theta}{1-\theta} e^{\left(x_{i}-z\right) b_{1}}}{1 + \frac{\theta}{1-\theta} e^{\left(x_{i}-z\right) b_{1}}} \prod_{j=1}^{n-r} \frac{1}{1 + \frac{\theta}{1-\theta} e^{\left(y_{j}-z\right) b_{1}}}.
\end{equation}
If we numerically integrate out the unwanted parameter $b_{1}$ out of \eqref{eq.1.12}, we get the posterior of the probability $\theta$, given the data $D$ and some predictor value $z$:
\begin{equation}
	\label{eq.1.13}
	p\!\left(\left.\theta\right|z, D\right) = \int p\!\left(\left.\theta, b_{1}\right|z, D\right) d b_{1},
\end{equation}
which gives us the Bayesian logistic regression model we are looking for. 

Note that for non-informative data, that is, for predictors which all have the same value, that is, $z = x_{i} = y_{j}$, for $i = 1, \dots, r$ and $j = 1, \dots, n-r$, the terms in the exponentials of \eqref{eq.1.12} all become zero, and, as a consequence,  \eqref{eq.1.13} collapses to the ordinary Beta distribution:
\begin{equation}
	\label{eq.1.14}
	p\!\left(\left.\theta\right|z, D\right) \propto	 \frac{1}{\theta\left(1-\theta\right)} \prod_{i=1}^{r} \frac{\frac{\theta}{1-\theta}}{1 + \frac{\theta}{1-\theta}} \prod_{j=1}^{n-r} \frac{1}{1 + \frac{\theta}{1-\theta}} = \theta^{r-1}\left(1-\theta\right)^{n-r-1},
\end{equation}
which is in nice correspondence with our intuition.

If the predictors are non-informative, in that they `flat-line', then the only pertinent aspect of the data $D$ which remains, is the number of successes and failures, respectively, $r$ and $n-r$, and these are just the sufficient statistics of the Beta distribution. 

\section{Times to Failure and Times Without Failure, the Exponential Model}

\subsection{The Probability Model.}
Say, we have an Exponential failure mechanism, then the probability of a failure at time $t$ is
\begin{equation}
	\label{eq.1.15}
	P\!\left(\left.t\right|\lambda\right) = \lambda \exp\left(-\lambda t\right) dt.
\end{equation}
Consequently, the probability of no failure until time $\tau$ is
\begin{equation}
	\label{eq.1.16}
	P\!\left(\left.t > \tau\right|\lambda\right) = \int_{\tau}^{\infty} \lambda \exp\left(-\lambda t\right) dt = \exp\left(-\lambda \tau \right).
\end{equation}

In most reliability problems we will be interested in determining probability \eqref{eq.1.16}. That is, in general we will wish to find the probability distribution of 
\begin{equation}
	\label{eq.1.17}
	\theta = \exp\left(-\lambda \tau \right),
\end{equation}
 where $\tau$ is some desired life-time and $\lambda$ is the unknown parameter of the Exponential distribution.
 
\subsection{The Likelihood, Prior, and Posterior.}
Say, we have $n$ identical units, which we follow in time. If we observe a sequence of $r$ failure times, say, $x_{1}, \ldots, x_{r}$, and $n-r$ units that did not fail, these having, consequently, having times without failure, say, $y_{1}, \ldots, y_{n-r}$. Then, by way of \eqref{eq.1.15} and \eqref{eq.1.16}, the probability of the observed data, or, equivalently, the likelihood of the unknown parameter $\lambda$, can be written down as
\begin{align}
	\label{eq.1.18}
	p\!\left(\left.D\right|\lambda\right) &= L\!\left(\lambda\right) \notag \\
	&= \prod_{i=1}^{r} \lambda \exp\left(-\lambda x_{i}\right) d x_{i} \prod_{j=1}^{n-r} \exp\left(-\lambda y_{j}\right) \\
	&\propto \prod_{i=1}^{r} \lambda \exp\left(-\lambda x_{i}\right) \prod_{j=1}^{n-r} \exp\left(-\lambda y_{j}\right), \notag
\end{align}   
where we let the constant term $\left(d x_{i}\right)^{r}$ be absorbed in the proportionality sign.

It would not be strange if our our prior information consisted of an initial gues of a life-time of, say, $t$. This initial guess corresponds with a prior likelihood of
\begin{equation}
	\label{eq.1.19}
	P\!\left(\left.t\right|\lambda\right) = \lambda \exp\left(-\lambda t \right) d t.
\end{equation}
Combining the prior likelihood \eqref{eq.1.19} with the uninformative Jeffreys' prior for the inverse failure rate $\lambda$,
\begin{equation}
	\label{eq.1.20}
	p\!\left(\lambda\right) \propto \frac{1}{\lambda},
\end{equation}
we get, by way of the product rule and the Bayesian proportionality short hand, the informative prior of $\lambda$, based on the initial guess of a life-time of $t$:
\begin{equation}
	\label{eq.1.21}
	p\!\left(\left.\lambda\right|t\right) \propto \exp\left(-\lambda t \right),
\end{equation}
 where we have absorbed both the differential $d t$ of \eqref{eq.1.19} and the normalizing constant of \eqref{eq.1.20} into the proportionality sign of \eqref{eq.1.21}.

Note that the prior \eqref{eq.1.21} may also be obtained through an alternative maximum entropy argument, \cite{Jaynes76}. But we give, instead, the above derivation. Because it is analogous to the derivation of the informative prior for a postulated Weibull failure mechanism, treated below.

Combining the likelihood with the informative prior, respectively, \eqref{eq.1.18} and \eqref{eq.1.21}, by way of the product rule and the Bayesian proportionality short hand, we get the posterior for the unknown parameter $\lambda$:
\begin{equation}
	\label{eq.1.22}
	p\!\left(\left.\lambda\right|D, t\right) \propto \exp\left(-\lambda t \right) 	\prod_{i=1}^{r} \lambda \exp\left(-\lambda x_{i}\right) \prod_{j=1}^{n-r} \exp\left(-\lambda y_{j}\right).
\end{equation}

The posterior \eqref{eq.1.22} is the probability distribution of the unknown parameter $\lambda$, conditional on the data $D$ we have observed and our tentative guess of a life-time of $t$. However, we are not that much interested in the probability distribution of $\lambda$. Rather, we are aiming for the probability distribution of the probability of the life-time exceeding $\tau$, \eqref{eq.1.17}.

Would we know the value of $\lambda$ exactly, then we could substitute this value into \eqref{eq.1.17} and, so, get the exact probability $\theta$. Now, we do not know the value of $\lambda$ exactly. Instead, we have a range of probable values on the $\lambda$-axis, as captured by the posterior \eqref{eq.1.22}. This corresponds, through a one-to-one mapping, with a range of probable values on the $\theta$-axis.

This one-to-one mapping is, typically, accomplished by way of a change of variable. 

\subsection{The Change of Variable.} 
By way of \eqref{eq.1.17}, we have that
\[
	\theta = \exp\left(-\lambda \tau \right).
\] 
So, the corresponding transformation is
\begin{equation}
	\label{eq.1.23}
	\lambda = -\frac{\log \theta}{\tau}, \qquad d \lambda = \left|-\frac{d \theta}{\theta \tau}\right| = \frac{d \theta}{\theta \tau}.
\end{equation}

Substituting the change of variable \eqref{eq.1.23} into the posterior \eqref{eq.1.22}, we obtain the transformed posterior distribution of the probability $\theta$ that a given unit will have a life-time exceeding $\tau$:
\begin{equation}
	\label{eq.1.24}
	p\!\left[\left.\theta\right|T\!\left(D,t\right), r, \tau\right] \propto \frac{\left(-\log \theta\right)^{r}}{\theta} \exp\left[\frac{\log \theta}{\tau} \; T\!\left(D,t\right)\right],
\end{equation}
where
\begin{equation}
	\label{eq.1.25}
	T\!\left(D,t\right) \equiv t + \sum_{i=1}^{r} x_{i} + \sum_{j=1}^{n-r} y_{j},
\end{equation}
is the total observed time without failure, $r$ is the number of observed failures, and $\tau$ is the life-time that has to be exceeded.

If we properly normalize \eqref{eq.1.24}, we get the identity:
\begin{equation}
	\label{eq.1.26}
	p\!\left[\left.\theta\right|T\!\left(D,t\right), r, \tau\right] = \left[\frac{T\!\left(D,t\right)}{\tau}\right]^{r+1} \frac{\left(-\log \theta\right)^{r}}{r ! \theta} \exp\left[\frac{\log \theta}{\tau} \; T\!\left(D,t\right)\right],
\end{equation}

By way of \eqref{eq.1.26}, the expectation value of the probability that a given unit will a life-time that exceeds $\tau$, then is
\begin{equation}
	\label{eq.1.27}
	E\!\left(\theta\right) = \int_{0}^{1} \theta \: p\!\left[\left.\theta\right|T\!\left(D,t\right), r, \tau\right] d \theta = \left[\frac{T\!\left(D,t\right)}{T\!\left(D,t\right) + \tau}\right]^{r+1}.
\end{equation}
This expectation value, which itself is a probability, is the result of Example~3, given in \cite{Jaynes76}. However, there it was not yet recognized that this probability is the mean of an underlying Beta-Like probability distribution\footnote{This, if anything, is an attestment to the richness of Jaynes' work. Even by working through the most casual of his derivations, one may still be rwarded for one's efforts by little gems, like the Beta-Like distributions given here.}

Following Jaynes, we subject \eqref{eq.1.27} to various extreme conditions, in order to show the correspondance with the indications of common sense. 

If the total, say, unit-hours of the test is small compared to prior expected life-time $t$, that is, if $\sum_{i} x_{i} + \sum_{j} y_{j} <\,< t$. Then, \eqref{eq.1.25},
\[
	T\!\left(D,t\right) \approx t,
\]
and, unless a large number of failures $r$ is observed, our state of knowledge about $\theta$ can hardly be changed by the test, and, as a consequence, we have to rely on our prior knowledge only.

If the total, say, unit-hours of the test is large compared to prior expected life-time $t$, that is, if $\sum_{i} x_{i} + \sum_{j} y_{j} >\,> t$. Then, \eqref{eq.1.25},
\[
	T\!\left(D,t\right) \approx \sum_{i=1}^{r} x_{i} + \sum_{j=1}^{n-r} y_{j},
\]
and, for all intents and purposes, our final conclusions depend only what we observed in the test, and, as a consequence, these conclusions are almost independent of what we previously thought previously.

In intermediate cases, our prior knowledge has a weight comparable to that of the test. If $t >\,>\tau$, the amount of testing required is appreciably reduced. For if we were already quite sure that the units are satisfactory, then we require less additional evidence before accepting them. But if $t <\,<\tau$, that is, if we are initially very doubtful about the units, then we demand that the test itself provide compelling evidence in favor of their reliability.

\section{Times to Failure Without Failure, the Weibull Model}
This is a repeat of the previous case, with the difference that we now use a Weibull failure mechanism instead of an Exponential one. 

\subsection{The Probability Model.}
Say, we have a Weibull failure mechanism, then the probability of a failure at time $t$ is
\begin{equation}
	\label{eq.1.28}
	P\!\left(\left.t\right|k, \lambda\right) = k \lambda \left(k \lambda\right)^{k-1} \exp\left[\left(- t \lambda\right)^{k}\right] d t.
\end{equation}
Consequently, the probability of no failure until time $\tau$ is
\begin{equation}
	\label{eq.1.29}
	P\!\left(\left.t > \tau \right|k, \lambda\right) = \int_{\tau}^{\infty} k \lambda \left(k \lambda\right)^{k-1} \exp\left[\left(- t \lambda\right)^{k}\right] d t =\exp\left[\left(- \tau \lambda\right)^{k}\right].
\end{equation}

In most reliability problems we will be interested in determining probability \eqref{eq.1.29}. That is, in general we wish to find the probability distribution of 
 \begin{equation}
	\label{eq.1.30}
	\theta = \exp\left[\left(-\lambda \tau \right)^{k}\right],
\end{equation}
where $\tau$ is some desired life-time, and $\lambda$ and $k$ are, respectively, the unknown inverse failure rate and shape parameter of the Weibull distribution.
 
\subsection{The Likelihood, Prior, and Posterior.}
Say, we have $n$ identical units, which we follow in time. If we observe a sequence of $r$ failure times, say, $x_{1}, \ldots, x_{r}$, and $n-r$ units that did not fail, these having, consequently, having times without failure, say, $y_{1}, \ldots, y_{n-r}$. Then, by way of \eqref{eq.1.28} and \eqref{eq.1.29}, the probability of the observed data, or, equivalently, the likelihood of the unknown parameters $\lambda$ and $k$, can be written down as
\begin{align}
	\label{eq.1.31}
	p\!\left(\left.D\right|k, \lambda\right) &= L\!\left(k, \lambda\right) \notag \\
	&= \prod_{i=1}^{r} k \lambda \left(\lambda x_{i}\right)^{k-1} \exp\left[\left(-\lambda x_{i}\right)^{k}\right] d x_{i} \prod_{j=1}^{n-r} \exp\left[\left(-\lambda y_{j}\right)^{k}\right] \\
	&\propto \prod_{i=1}^{r} k \lambda \left(\lambda x_{i}\right)^{k-1} \exp\left[\left(-\lambda x_{i}\right)^{k}\right] \prod_{j=1}^{n-r} \exp\left[\left(-\lambda y_{j}\right)^{k}\right] , \notag
\end{align}   
where we let the constant term $\left(d x_{i}\right)^{r}$ be absorbed in the proportionality sign.

If our our prior information consisted of an initial gues of a life-time of, say, $t$. This initial guess corresponds with a prior likelihood of
\begin{equation}
	\label{eq.1.32}
	P\!\left(\left.t\right|k, \lambda\right) = k \lambda \left(\lambda t\right)^{k-1} \exp\left[\left(-\lambda t \right)^{k}\right] d t.
\end{equation}
Combining the prior likelihood \eqref{eq.1.32} with the uninformative Jeffreys' prior for the inverse failure rate and shape parameter, respectively, $\lambda$ and $k$,
\begin{equation}
	\label{eq.1.33}
	p\!\left(k, \lambda\right) = p\!\left(k\right) p\!\left(\lambda\right) \propto \frac{1}{k \lambda},
\end{equation}
we get, by way of the product rule and the Bayesian proportionality short hand, the informative prior of $\lambda$ and $k$, based on the initial guess of a life-time of $t$:
\begin{equation}
	\label{eq.1.34}
	p\!\left(\left.k, \lambda\right|t\right) \propto \left(\lambda t\right)^{k-1} \exp\left[\left(-\lambda t \right)^{k}\right],
\end{equation}
 where we have absorbed both the differential $d t$ of \eqref{eq.1.32} and the normalizing constant of \eqref{eq.1.33} into the proportionality sign of \eqref{eq.1.34}.

In \cite{vanErp11}, an alternative informative prior is derived, where the piece of prior information consists of an initial guess of the time without failure, $s^{*}$. Now, would we have no initial guess whatsoever, neither for the time to failure nor for the time without failure, then the proper cause of action would be to assign as an uninformative prior the prior \eqref{eq.1.33}.

Combining the likelihood with the informative prior, respectively, \eqref{eq.1.31} and \eqref{eq.1.34}, by way of the product rule and the Bayesian proportionality short hand, we get the posterior for the unknown parameters $\lambda$ and $k$:
\begin{equation}
	\label{eq.1.35}
	p\!\left(\left.k, \lambda\right|D, t\right) \propto \left(\lambda t\right)^{k-1} \exp\left[\left(-\lambda t \right)^{k}\right] 	\prod_{i=1}^{r} k \lambda \left(\lambda x_{i}\right)^{k-1} \exp\left[\left(-\lambda x_{i}\right)^{k}\right] \prod_{j=1}^{n-r} \exp\left[\left(-\lambda y_{j}\right)^{k}\right].
\end{equation}

The posterior \eqref{eq.1.35} is the probability distribution of the unknown parameters $\lambda$ and $k$, conditional on the data $D$ we have observed and our tentative guess of a life-time of $t$. However, we are not that much interested in the probability distribution of $\lambda$ and $k$. Rather, we are aiming for the probability distribution of the probability of the life-time exceeding $\tau$, \eqref{eq.1.30}.

Would we know the values of $\lambda$ and $k$ exactly, then we could substitute this value into \eqref{eq.1.30} and, so, get the exact probability $\theta$. Now, we do not know the values of $\lambda$ and $k$ exactly. Instead, we have a range of probable values on the $\lambda$- and $k$-axes, as captured by the posterior \eqref{eq.1.35}. This corresponds, through a two-to-one mapping, with a range of probable values on the $\theta$-axis.

This two-to-one mapping is, typically, accomplished by way of a Jacobian transformation. 
 
\subsection{The Jacobian Transformation.}
By way of \eqref{eq.1.30}, we have that
\[
	\theta = \exp\left[\left(-\lambda \tau \right)^{k}\right].
\] 
So, a possible transformation is
\begin{equation}
	\label{eq.1.36}
	\lambda = \frac{\left(-\log \theta\right)^{\frac{1}{\kappa}}}{\tau}, \qquad k = \kappa,
\end{equation}
which has a corresponding Jacobian of
\begin{equation}
	\label{eq.1.37}
	J = 
\left|\begin{array}{cc}
	\frac{\partial}{\partial \theta} \lambda & \frac{\partial}{\partial \kappa} \lambda \\
	\frac{\partial}{\partial \theta} k & \frac{\partial}{\partial \kappa} k 
\end{array}\right| = \left|\begin{array}{cc}
	- \frac{\left(-\log \theta\right)^{\frac{\kappa-1}{\kappa}}}{\tau \kappa \theta} & f\!\left(\theta, \kappa\right) \\
	0 & 1
\end{array}\right| \propto   \frac{\left(-\log \theta\right)^{-\frac{\kappa-1}{\kappa}}}{\tau \kappa \theta}.
\end{equation}


Substituting \eqref{eq.1.36} into the posterior \eqref{eq.1.35}, and multiplying it with the Jacobian \eqref{eq.1.37}, gives us the transformed posterior:
\begin{equation}
	\label{eq.1.38}
	p\!\left[\left.\kappa, \theta\right| T\!\left(D, t, \kappa\right), r, \tau\right] \propto \left(t \prod_{i=1}^{r} x_{i}\right)^{\kappa-1} \frac{\kappa^{r-1}}{\tau^{\kappa \left(r + 1\right)}} \frac{\left(-\log \theta\right)^{r}}{\theta} \exp\left[\frac{\log\theta}{\tau^{\kappa}} T\!\left(D, t, \kappa\right)\right],
\end{equation}
where $\kappa$ is the shape parameter of the Weibull distribution and
where
\begin{equation}
	\label{eq.1.39}
	T\!\left(D,t,\kappa\right) \equiv t^{\kappa} + \sum_{i=1}^{r} x_{i}^{\kappa} + \sum_{j=1}^{n-r} y_{j}^{\kappa},
\end{equation}
is the total power-transformed observed time without failure, $r$ is the number of observed failures, and $\tau$ is the life-time that has to be exceeded.

Note that if we set the shape parameter to $\kappa = 1$, or, equivalently, we go from the Weibull to the more restrictive Exponential failure mechanism, then, by way of the proportionality sign, the posterior \eqref{eq.1.38} collapses to \eqref{eq.1.24}, which is at should be.

Looking at the statistic \eqref{eq.1.39}, we see that for, say, $\kappa = 2$, one observation having no failure until time $y = 7$ is equivalent to 49 observations having no failure until time $y = 1$.

For $\kappa = 1$, where the Weibull collapses to the memoryless Exponential distribution, one observation having no failure time until time $y = 7$ is equivalent to 7 observations having no failure until time $y = 1$. 

This reflects the Weibull's dependence on the shape parameter $\kappa$. For large values of $\kappa$, extended periods without failure become less probable. Thus, observing one extended period without failure carries the same weight as observing many more short periods without failure.

If we numerically integrate out the unwanted parameter $\kappa$ out of \eqref{eq.1.38}, we get the posterior of the probability $\theta$, given the data $D$ and the initial guess of time without failure, $t$:
\begin{equation}
	\label{eq.1.40}
	p\!\left(\left.\theta\right| t, D, \tau\right) = \int p\!\left[\left.\kappa, \theta\right| T\!\left(D, t, \kappa\right), r, \tau\right] d \kappa.
\end{equation}

\chapter{Implicit Probability Distributions for Probabilities, Part I}
There are instances were we cannot rewrite any of the unknown parameters in the posterior as a function of $\theta$, as was done, for example, in \eqref{eq.1.10}, \eqref{eq.1.23}, and \eqref{eq.1.36}. This, then, prohibits us from finding the explicit form of the corresponding Beta-like distribution.

However, we may still find the first moments of these intractable distributions. Thus, allowing us to either approximate the corresponding probability distribution or, alternatively, to construct confidence bounds on the estimated probabilities.

\section{Times to $m$ Events, the Poisson Model}

\subsection{The Probability Model.}
Here we define the probability of interest, $\theta$, to be the probability of $m$ events occurring within the time period $\tau$, by a mechanism which is modeled by an underlying Exponential distribution, having an unknown parameter $\lambda$. 

The data consists of single events observed at variable, though consecutive, waiting times $t_{1},\dots,t_{m}$, and a non-event from the last failure, which is observed at time $\sum_{j} t_{j}$, onwards, until the end of the time period $\tau$:
\begin{align}
	\label{eq.2.1}
	P\!\left(\left.m\right| \lambda, \tau\right) &= \int_{0}^{\tau} \int_{0}^{\tau - t_{1}} \cdots \int_{0}^{\tau - \sum_{j=1}^{m-1}t_{j}} \lambda e^{- \lambda t_{1}} \lambda e^{- \lambda t_{2}} \cdots \lambda e^{- \lambda t_{m}} e^{- \lambda \left(\tau - \sum_{j} t_{j}\right)} dt_{m}\cdots dt_{2} dt_{1}, \notag \\
	&= \frac{\left(\lambda \tau\right)^{m}}{m!} \exp\left(- \lambda \tau\right).
\end{align}

An inspection of \eqref{eq.2.1} learns us that the probability of interest has the form of a Poisson distribution, which has an expected number of events equal to $\lambda \tau$. So, we define our probability of interest to be
\begin{equation}
	\label{eq.2.2}
	\theta = \frac{\left(\lambda \tau\right)^{m}}{m!} \exp\left(- \lambda \tau\right).
\end{equation}

\subsection{The Likelihood, Prior, and Posterior.}
In a total time period of, say, $T$, we observe $n$ consecutive times to an even, $x_{1}, \ldots, x_{n}$. Assuming an Exponential event-generating mechanism,
\begin{equation}
	\label{eq.2.3}
	P\!\left(\left.x_{i}\right|\lambda\right) = \lambda \exp\left(-\lambda x_{i}\right) dx_{i},
\end{equation} 
the probability of the observed data, or, equivalently, the likelihood of the unknown parameter $\lambda$, which is the expected of events per time unit, may be written down as
\begin{align}
	\label{eq.2.4}
	P\!\left(\left.D\right|\lambda\right) &= L\!\left(\lambda\right) \notag \\
	&= \exp\left[-\lambda\left(T-\sum_{i} x_{i}\right)\right] \prod_{i=1}^{n} \lambda \exp\left(-\lambda x_{i}\right) dx_{i} \\
	& \propto \lambda^{n} \exp\left(-\lambda T\right), \notag
\end{align} 
where we have absorbed the term $\left(dx_{i}\right)^{n}$ into the proportionality sign.

If we have an initial guess that the time to an event is $t$, we may assign the informative prior \eqref{eq.1.21}
\begin{equation}
	\label{eq.2.5}
	p\!\left(\left.\lambda\right|t\right)\propto \exp\left(-\lambda t\right).
\end{equation} 
However, if we do not feel confident enough, to make such a prior guess, we may alternatively, assign an uninformative Jeffreys prior
\[
	p\!\left(\lambda\right)\propto \frac{1}{\lambda}.
\]

Multiplying the likelihood with the informative prior, respectively, \eqref{eq.2.4} and \eqref{eq.2.5}, we may obtain the posterior of $\lambda$:
\begin{equation}
	\label{eq.2.6}
	p\!\left(\left.\lambda\right|D, t\right) \propto \lambda^{n} \exp\left[-\lambda \left(T + t\right)\right].
\end{equation} 
where the pertinent aspects of the data $D$ are the number of events, $n$, and the total time of observation $T$.

As an aside, if we have two data sets of the same phenomena under observation, say, $D_{1}$ and $D_{2}$, having, respectively, $n_{1}$ and $n_{2}$ observed events in the respective periods $T_{1}$ and $T_{2}$, then these data sets, together with the informative prior \eqref{eq.2.5}, would combine to the posterior
\[
	p\!\left(\left.\lambda\right|D_{1}, D_{2}, t\right) \propto \lambda^{n_{1} + n_{2}} \exp\left[-\lambda \left(T_{1} + T_{2} + t\right)\right].
\]
 
Now, would we attempt in \eqref{eq.2.2} to make a change of variable from $\lambda$ to $\theta$, then we find that $\lambda$ cannot be written unambiguously as a function of $\theta$ for $m > 0$; where $m = 0$ is equivalent to the Exponential probability model, \eqref{eq.1.17}. It follows that we can make no analytical change of variable for the Poission model probability \eqref{eq.2.2}, or, equivalently, derive the explicit Beta-Like distribution for this probability model.

But what we can do, is derive the first moments of this intractable Beta-Like distribution. This will allow us to either compute the skewness corrected confidence bounds for this intractable distribution, or, alternatively, construct the MaxEnt distribution which has the same moments as this intractable distribution, respectively, \cite{vanErp14} and \cite{Rockinger01}.

\subsection{MaxEnt Distributions}
\label{MaxEnt} 
In what follows we give a short outline on how to derive fourth-order MaxEnt distributions. Three well-known MaxEnt distributions are the uniform, exponential, and normal distributions. These distributions correspond, respectively, with zeroth-, first-, and second-order MaxEnt distributions. 

For a given probability distribution
\[
	p\!\left(\left.\theta\right|\left\{\boldsymbol{\lambda}\right\}\right),
\]
where $\theta$ is the parameter of interest and $\left\{\boldsymbol{\lambda}\right\}$ is some set of parameters which make up this probability distribution, the first four cumulants are given as:
\begin{align}
	\label{eq.2.7}
	\mu &= \int \theta \: p\!\left(\left.\theta\right|\left\{\boldsymbol{\lambda}\right\}\right) d\theta, \notag \\
	\sigma &= \sqrt{\int \left(\theta - \mu\right)^{2} \: p\!\left(\left.\theta\right|\left\{\boldsymbol{\lambda}\right\}\right) d\theta}, \notag \\
	\gamma &= \frac{\int \left(\theta - \mu\right)^{3} \: p\!\left(\left.\theta\right|\left\{\boldsymbol{\lambda}\right\}\right) d\theta}{\sigma^{3}}, \\
	\kappa &= \frac{\int \left(\theta - \mu\right)^{4} \: p\!\left(\left.\theta\right|\left\{\boldsymbol{\lambda}\right\}\right) d\theta}{\sigma^{4}}. \notag 
\end{align}
where $\mu$ is the mean, $\sigma$ is the standard deviation, $\gamma$ is the skewness, and $\kappa$ is the kurtosis of the probability distribution $p\!\left(\left.\theta\right|\left\{\boldsymbol{\lambda}\right\}\right)$.

The fourth-order MaxEnt distribution incorporates information about the skewness and kurtosis, respectively, $\gamma$ and $\kappa$, \eqref{eq.2.7}, as well as the mean and standard deviation, respectively, $\mu$ and $\sigma$. The algorithm for higher-order MaxEnt distributions is due to Rockinger and Jondeau, \cite{Rockinger01}. 

We will now proceed to give the algorithmic steps needed to construct fourth-order MaxEnt approximations of intractable Beta-like distributions.\\
\\
\noindent\textbf{Step 1.} \\
\noindent The fourth-order MaxEnt distribution we seek takes as its input the first four cumulants of the probability distribution we wish to approximate. 

For example, if wish to determine the fourth-order MaxEnt distrbution of $\theta$. Then we first compute the first four moments of $\theta$, \eqref{eq.2.2}, \eqref{eq.2.6}, and \eqref{eq.2.7}:
\begin{align}
	\label{eq.2.11}
	m_{1} &= \int \left[\frac{\left(\lambda \tau\right)^{m}}{m!} \exp\left(- \lambda \tau\right)\right]^{1}  \lambda^{n} \exp\left[-\lambda \left(T + t\right)\right] d\lambda,  \notag \\
	m_{2} &= \int \left[\frac{\left(\lambda \tau\right)^{m}}{m!} \exp\left(- \lambda \tau\right)\right]^{2}  \lambda^{n} \exp\left[-\lambda \left(T + t\right)\right] d\lambda, \notag \\
	m_{3} &= \int \left[\frac{\left(\lambda \tau\right)^{m}}{m!} \exp\left(- \lambda \tau\right) \right]^{3}  \lambda^{n} \exp\left[-\lambda \left(T + t\right)\right] d\lambda, \\
	m_{4} &= \int \left[\frac{\left(\lambda \tau\right)^{m}}{m!} \exp\left(- \lambda \tau\right) \right]^{4} \lambda^{n} \exp\left[-\lambda \left(T + t\right)\right] d\lambda, \notag 
\end{align}

The moments in \eqref{eq.2.11} evaluate to
\begin{align}
	\label{eq.2.12}
	m_{1} &= \frac{\left(m + n\right)!}{m! n!} \left(\frac{\tau}{\tau + T + t}\right)^{m} \left(\frac{T + t}{\tau + T + t}\right)^{n+1}, \notag \\
	m_{2} &= \frac{\left(2 m + n\right)!}{\left(m!\right)^{2} n!} \left(\frac{\tau}{2 \tau + T + t}\right)^{2 m} \left(\frac{T + t}{2 \tau + T + t}\right)^{n+1}, \notag \\
	m_{3} &= \frac{\left(3 m + n\right)!}{\left(m!\right)^{3} n!} \left(\frac{\tau}{3 \tau + T + t}\right)^{3 m} \left(\frac{T + t}{3 \tau + T + t}\right)^{n+1}, \\
	m_{4} &= \frac{\left(4 m + n\right)!}{\left(m!\right)^{4} n!} \left(\frac{\tau}{4 \tau + T + t}\right)^{4 m} \left(\frac{T + t}{4 \tau + T + t}\right)^{n+1}. \notag 
\end{align}

By way of \eqref{eq.2.12} and the identities, \cite{Hall92},
\begin{align}
	\label{eq.2.13}
	\mu &= m_{1},   \notag \\
	\sigma &= \sqrt{m_{2} - m_{1}^{2}}, \notag \\
	\gamma &= \frac{m_{3} - 3 m_{2} \: m_{1} + 2 m_{1}^{3}}{\sigma^{3}}, \\
	\kappa &= \frac{m_{4} - 4 m_{3} \: m_{1} + 6 m_{2} \: m_{1}^{2} - 3 m_{1}^{4}}{\sigma^{4}}, \notag 
\end{align}
we may then compute the first four cumulants, needed for the construction of the fourth-order MaxEnt distribution.\\
\\
\noindent\textbf{Step 2.} \\
\noindent We now plug the third and fourth cumulant, respectively, $\gamma$ and $\kappa$, into the integral function
\begin{equation}
	\label{eq.2.14}
	Q\!\left(\varphi_{1}, \varphi_{2}, \varphi_{3}, \varphi_{4}\right) = \int_{a}^{b} \exp\left[\varphi_{1} x + \varphi_{2} \left(x^{2}-1\right) + \varphi_{3} \left(x^{3}-\gamma\right) + \varphi_{4} \left(x^{4}-\kappa\right)\right] dx,
\end{equation}
where\footnote{The limits of integration are determined by the identities:
\[
	\mu + a \sigma = 0, \qquad \mu + b \sigma = 1.
\]}
\begin{equation}
	\label{eq.2.15}
	a = -\frac{\mu}{\sigma}, \qquad  b = \frac{1 - \mu}{\sigma}.
\end{equation}

Then minimize \eqref{eq.2.14} over the vector $\left(\varphi_{1}, \varphi_{2}, \varphi_{3}, \varphi_{4}\right)$, in order to obtain the minimization estimates $\left(\hat{\varphi}_{1}, \hat{\varphi}_{2}, \hat{\varphi}_{3}, \hat{\varphi}_{4}\right)$. \\
\\
\noindent\textbf{Step 3.} \\
\noindent We then make a change of variable from $x$ to $\theta$, where
\begin{equation}
	\label{eq.2.16}
	x = \frac{\theta - \mu}{\sigma},
\end{equation}
in order to obtain the unscaled MaxEnt distribution on the $\theta$ axis:
\begin{equation}
	\label{eq.2.17}
	p\!\left(\left.\theta \right| \mu,\sigma,\gamma,\kappa\right) \propto \frac{1}{\sigma} \exp\left[\hat{\varphi}_{1} \frac{\theta - \mu}{\sigma} + \hat{\varphi}_{2} \left(\frac{\theta - \mu}{\sigma}\right)^{2} + \hat{\varphi}_{3} \left(\frac{\theta - \mu}{\sigma}\right)^{3} + \hat{\varphi}_{4} \left(\frac{\theta - \mu}{\sigma}\right)^{4}\right].
\end{equation}
The normalizing constant of \eqref{eq.2.17} then may be computed by way of the integral, \eqref{eq.2.15} and \eqref{eq.2.16}, 
\begin{equation}
	\label{eq.2.18}
	C = \int_{0}^{1} p\!\left(\left.\theta\right|\mu,\sigma,\gamma,\kappa\right) d\theta, 
\end{equation}

The properly normalized fourth-order MaxEnt approximation of the intractable beta-like distribution, which has its probability model \eqref{eq.2.2}, is then given as, \eqref{eq.2.12}, \eqref{eq.2.13}, \eqref{eq.2.14}, \eqref{eq.2.17}, and \eqref{eq.2.18}:
\begin{equation}
	\label{eq.2.19}
	p\!\left(\left.\theta\right|\mu,\sigma,\gamma,\kappa\right) = \frac{1}{C \sigma} \exp\left[\hat{\varphi}_{1} \frac{\theta - \mu}{\sigma} + \hat{\varphi}_{2} \left(\frac{\theta - \mu}{\sigma}\right)^{2} + \hat{\varphi}_{3} \left(\frac{\theta - \mu}{\sigma}\right)^{3} + \hat{\varphi}_{4} \left(\frac{\theta - \mu}{\sigma}\right)^{4}\right],
\end{equation}
where $0 \leq \theta \leq 1$.

\section{Times to $m$ Events, the Cumulative Poisson Model}

\subsection{The Probability Model.}
In the preceeding discussion we discussed the Beta-Like distribution of the probability of $m$ events occurring within the time period $\tau$, by a mechanism which is modeled by an underlying Exponential distribution, having an unknown parameter $\lambda$. This probability distribution may be of interest if we have a parallel system of $m$ non-redundent fail-safe mechanisms, where each mechanism admits an Exponential time to failure model.

Now, we can envisage scenarios in which we wish to determine the Beta-Like distribution of the probability of more than $m$ events occurring within the time period $\tau$, by a mechanism which is modeled by an underlying Exponential distribution, having an unknown parameter $\lambda$. One such scenario may be where we have a system which is subject to successive loads, each load admitting an Exponential time to occurrence model. The system then might be hypothesized to be able up to $m$ such loads, before a significant wear and tear occurs.

For this scenario the probability model of interest is
\begin{equation}
	\label{eq.2.20}
	\theta = 1 - \sum_{i=0}^{m} \frac{\left(\lambda \tau\right)^{i}}{i!} \exp\left(- \lambda \tau\right).
\end{equation}

Now, \eqref{eq.2.20} is just a probability, just like, say, \eqref{eq.2.2} is, which admits an uncertainty regarding the actual value of the inverse failure rate $\lambda$, as captured by the posterior \eqref{eq.2.6}. So, we may proceed to compute the first four moments of the Beta-Like probability distribution which results from the uncertainty we have regarding the actual value of $\lambda$:
\begin{align}
	\label{eq.2.21}
	m_{1} &= \int \left[1 - \sum_{i=0}^{m} \frac{\left(\lambda \tau\right)^{i}}{i!} \exp\left(- \lambda \tau\right)\right]^{1}  \lambda^{n} \exp\left[-\lambda \left(T + t\right)\right] d\lambda,  \notag \\
	m_{2} &= \int \left[1 - \sum_{i=0}^{m} \frac{\left(\lambda \tau\right)^{i}}{i!} \exp\left(- \lambda \tau\right)\right]^{2}  \lambda^{n} \exp\left[-\lambda \left(T + t\right)\right] d\lambda, \notag \\
	m_{3} &= \int \left[1 - \sum_{i=0}^{m} \frac{\left(\lambda \tau\right)^{i}}{i!} \exp\left(- \lambda \tau\right)\right]^{3}  \lambda^{n} \exp\left[-\lambda \left(T + t\right)\right] d\lambda, \\
	m_{4} &= \int \left[1 - \sum_{i=0}^{m} \frac{\left(\lambda \tau\right)^{i}}{i!} \exp\left(- \lambda \tau\right)\right]^{4} \lambda^{n} \exp\left[-\lambda \left(T + t\right)\right] d\lambda, \notag 
\end{align}
Having obtained these moments we may compute the relevant cumulants, by way of \eqref{eq.2.13}, and either proceed to construct the fourth-order MaxEnt approximation of the intractable Beta-Like distribution of \eqref{eq.2.19}. 

\section{Predictors, the Poisson Regression Model}

\subsection{The Probability Model.}
In the Poisson regression model the number of events occuring, $m$, in a given time period, $\tau$, has Poisson distribution:
\begin{equation}
	\label{eq.2.22} 
	P\!\left(\left.m\right| \lambda, \tau \right) = \frac{\left(\lambda \tau\right)^{m}}{m!} \exp\left(- \lambda \tau\right),
\end{equation}
where the logarithm of the expected number of events per time unit, $\lambda$, is modeled by way of a regression model:
\begin{equation}
	\label{eq.2.23} 
	\log \lambda = \beta_{0} + \beta_{1} z.
\end{equation}
If we take the exponential of \eqref{eq.2.23}, and substitute it into \eqref{eq.2.22}, we get the probability model:
\begin{align}
	\label{eq.2.24} 
	\theta &= \frac{\left[\exp\left(\beta_{0} + \beta_{1} z\right) \tau\right]^{m}}{m!} \exp\left[- \exp\left(\beta_{0} + \beta_{1} z\right) \tau\right] \notag \\ 
	&= \frac{\tau^{m}}{m!} \exp\left[m \left(\beta_{0} + \beta_{1} z\right) - \exp\left(\beta_{0} + \beta_{1} z\right) \tau\right]
\end{align}

\subsection{The Likelihood, Prior, and Posterior.}
The data $D$ consists of $n$ counts, $r_{1}, \dots, r_{n}$, with corresponding predictor values $x_{1}, \dots, x_{n}$.  Using \eqref{eq.2.24}, the probability of the data $D$, or, equivalently, the likelihood of the unknown parameters $\beta_{0}$ and $\beta_{1}$, may be written down as
\begin{align} 
	\label{eq.2.25} 
	p\!\left(\left.D\right|\beta_{0}, \beta_{1}\right) &= \prod_{i=1}^{n} \frac{\tau^{r_{i}}}{r_{i}!} \exp\left[r_{i} \left(\beta_{0} + \beta_{1} x_{i}\right) - \exp\left(\beta_{0} + \beta_{1} x_{i}\right) \tau\right] \notag\\
	&\propto \prod_{i=1}^{n} \exp\left[r_{i} \left(\beta_{0} + \beta_{1} x_{i}\right) - \exp\left(\beta_{0} + \beta_{1} x_{i}\right) \tau\right] \\
	&\propto \exp\left[\sum_{i} r_{i} \left(\beta_{0} + \beta_{1} x_{i}\right) - \tau \: \sum_{i} \exp\left(\beta_{0} + \beta_{1} x_{i}\right) \right] \notag
\end{align}

We assign some uniform prior to the unknown regression parameters $\beta_{0}$ and $\beta_{1}$:
\begin{equation} 
	\label{eq.2.26} 
	p\!\left(\left.D\right|\beta_{0}, \beta_{1}\right) \propto \prod_{i=1}^{n} \exp\left[r_{i} \left(\beta_{0} + \beta_{1} x_{i}\right) - \exp\left(\beta_{0} + \beta_{1} x_{i}\right) \tau\right]
\end{equation}

By multiplying the likelihood \eqref{eq.2.25} with the prior \eqref{eq.2.26}, one may obtain the unscaled posterior of $\beta_{0}$ and $\beta_{1}$:
\begin{equation} 
	\label{eq.2.27} 
		p\!\left(\left.\beta_{0}, \beta_{1}\right|D\right) \propto p\!\left(\beta_{0}, \beta_{1}\right) \propto \exp\left[\sum_{i} r_{i} \left(\beta_{0} + \beta_{1} x_{i}\right) - \tau \: \sum_{i} \exp\left(\beta_{0} + \beta_{1} x_{i}\right) \right].
\end{equation}

Since we can make no analytical Jacobian transformation from the $\beta_{0}$ and $\beta_{1}$ to the probability model $\theta$, we compute the first moments of \eqref{eq.2.24}, by way of the posterior \eqref{eq.2.27}, and proceed to construct either the skewness corrected confidence interval or the MaxEnt approximation of the corresponding Beta-Like distribution.  

\chapter{Implicit Probability Distributions for Probabilities, Part II}
We will here construct the Beta-Like distribution of a Poisson-Like probability model. We define the probability of interest, $\theta$, to be the probability of $m$ events occurring within the time period $\tau$, by a mechanism which is modeled by an underlying Weibull distribution, having an unknown parameters $\lambda$ and $k$.

The advantage of a Weibull over an Exponential mechanism, is that the shape parameter $k$ of the former represents an extra degree of freedom, as it may take on any value greater than zero; whereas, in the latter it is dogmatically set to one. 

Regular Poisson distributions have as their event-generating mechanism Exponential distributions, \eqref{eq.2.1}. In contrast, a sequence of Weibull distributed events leaves us with an analytically intractable integral. 

However, making use of an unexpected equivalence relationship, we may work around the encountered integral and, so, proceed to approximate the Poisson-Like distribution.

\section{The Issue}
We first present the case for the regular Poisson distribution, as this will give us a handle on how to generalize from the Poisson to the Poisson-Like distribution.

Let $\theta$ be the probability of $m$ failures occurring within time period $\tau$. The failure mechanism generating these events is the Exponential distribution, having an unknown parameter $\lambda$. The data we will use is failures at variable, though consecutive, times $t_{1}, \ldots, t_{m}$, and a non-failure from the last failure, at time $t_{m}$, onwards, until the end of the time period $\tau$, \eqref{eq.2.1}:
\begin{align}
	\label{eq.3.1}
	\theta &= \int_{0}^{\tau} \int_{0}^{\tau - t_{1}} \cdots \int_{0}^{\tau - \sum_{j=1}^{m-1}t_{j}} \lambda e^{- \lambda t_{1}} \lambda e^{- \lambda t_{2}} \cdots \lambda e^{- \lambda t_{m}} e^{- \lambda \left(\tau - \sum_{j} t_{j}\right)} dt_{m}\cdots dt_{2} dt_{1}, \notag \\
	&= \lambda^{m} e^{- \lambda \tau} \int_{0}^{\tau} \int_{0}^{\tau - t_{1}} \cdots \int_{0}^{\tau - \sum_{j=1}^{m-1}t_{j}} dt_{m}\cdots dt_{2} dt_{1},
\end{align}
where it may be found, by way of induction, that
\begin{equation}
	\label{eq.3.2}
	  \int_{0}^{\tau} \int_{0}^{\tau - t_{1}} \cdots \int_{0}^{\tau - \sum_{j=1}^{m-1}t_{j}}  dt_{m}\cdots dt_{2} dt_{1} = \frac{\tau^{m}}{m!}.
\end{equation}

Substituting \eqref{eq.3.2} into \eqref{eq.3.1}, we find that $\theta$ is just the traditional Poisson probability of $m$ events occurring in a time period $\tau$, \eqref{eq.2.2}:
\begin{equation}
	\label{eq.3.3}
	 \theta = \frac{\left(\lambda \tau\right)^{m}}{m!} \exp\left(- \lambda \tau\right),
\end{equation}
where $\lambda \tau$ is the expected number of failures within the time period $\tau$.

We again define $\theta$ to be the probability of $m$ failures occurring within the time period $\tau$, but now by a failure mechanism which is modelled by an unerlying Weibull distribution having parameters $\lambda$ and $k$.

The Weibull model, having one more parameter, is more flexible than the Exponential model. In fact, the Exponential is a special case of the Weibull, were we set the shape paramter $k$ to one. 

The data, again, consists of failures at variable, though consecutive, times $t_{1}, \dots, t_{m}$, and a non-failure from the last failure, at time $t_{m}$, onwards, until the end of the time period $\tau$:
\begin{align}
	\label{eq.3.4}
	\theta = \int_{0}^{\tau} \int_{0}^{\tau - t_{1}} \cdots \int_{0}^{\tau - \sum_{j=1}^{m-1}t_{j}} &k \lambda \left(\lambda t_{1}\right)^{k-1} e^{- \left(\lambda t_{1}\right)^{k}} k \lambda \left(\lambda t_{2}\right)^{k-1} e^{- \left(\lambda t_{2}\right)^{k}} \cdots \\
	&\cdots k \lambda \left(\lambda t_{m}\right)^{k-1} e^{- \left(\lambda t_{m}\right)^{k}} e^{- \left[\lambda \left(\tau - \sum_{j} t_{j}\right)\right]^{k}} dt_{m}\cdots dt_{2} dt_{1}. \notag
\end{align}

Integral \eqref{eq.3.4} does not allow for a simple analytical expression like \eqref{eq.3.1}. So, by way of the curse of dimensionality, as $m >\!> 1$, we are prohibited from evaluating the first cumulants of \eqref{eq.3.4} and, as a consequence, constructing either a confidence bound or a MaxEnt approximative distribution. However, there is a useful equivalence relation which may be used to find these cumulants after all.

\section{The Equivalence Relation}
The equivalence relationship is derived first for the regular Poisson model \eqref{eq.3.1}. Since, only for this regular case do we have the analytical solution of the target integral with which we can compare the alternative route.

Let $q$ be the sum of $n + 1$ waiting times, which are generated by independent Exponential processes:
\begin{equation}
	\label{eq.3.5}
	q = t_{1} + \cdots + t_{n+1}, \qquad t_{i}  \thicksim Exp\left(\lambda\right).
\end{equation}
Then we may derive the probability density function of the stochastic $q$ from the product of $n + 1$ Exponential distributions,
\begin{equation}
	\label{eq.3.6}
	p\!\left(\left.t_{1},\ldots, t_{n+1}\right|\lambda\right) = \prod_{i=1}^{n+1} \lambda \exp\left(\lambda t_{i}\right),
\end{equation}
and some appropriate Jacobian transformation like, for example,
\begin{equation}
	\label{eq.3.7}
	\begin{cases}
	t_{1} = q - t_{2}' - \cdots - t_{n + 1}',  \\
	t_{2} = t_{2}', \\
	\quad \vdots \\
	t_{n+1} = t_{n+1}'.
	\end{cases}
\end{equation}

By way of \eqref{eq.3.6} and \eqref{eq.3.7}, while keeping track of the appropriate integration limits of the $n$ unwanted parameters, $t_{2}', \ldots, t_{n+1}'$, we may find the probability distribution of $q$:
\begin{align}
	\label{eq.3.8}
	p\!\left(\left.q\right|\lambda\right) &= \lambda^{n+1} \exp\left(-\lambda q\right) \int_{0}^{q} \int_{0}^{q - t_{2}'} \cdots \int_{0}^{q - \sum_{i=2}^{n+1} t_{i}'} dt_{m}'\cdots  dt_{3}' dt_{2}' \notag \\
	&= \lambda \frac{\left(\lambda q\right)^{n}}{n!} \exp\left(-\lambda q\right).
\end{align} 

Now, as it turns out, the cumulative distribution function of $q$, that is, the sum of $n + 1$ Exponential waiting times,
\begin{equation}
	\label{eq.3.9}
	P\!\left(\left.q \leq \tau \right| \lambda \right) = \int_{0}^{\tau} 	\lambda \frac{\left(\lambda q\right)^{n}}{n!} \exp\left(-\lambda q\right) dq,
\end{equation}
is equivalent to the probability of observing more than $n$ events in the time period $\tau$, 
\begin{equation}
	\label{eq.3.10}
	P\!\left(\left.i > n \right| \lambda, \tau\right) = 1 - \sum_{i=0}^{n} \frac{\left(\lambda \tau\right)^{i}}{i!} \exp\left(- \lambda \tau\right).
\end{equation}
So, the equivalence relationship we will make use of is, \eqref{eq.3.9} and \eqref{eq.3.10}, see also \eqref{eq.2.20}, 
\begin{equation}
	\label{eq.3.11}
	\int_{0}^{\tau} 	\lambda \frac{\left(\lambda q\right)^{n}}{n!} \exp\left(-\lambda q\right) dq = 1 - \sum_{i=0}^{n} \frac{\left(\lambda \tau\right)^{i}}{i!} \exp\left(- \lambda \tau\right),
\end{equation}  
or, equivalently, 
\begin{equation}
	\label{eq.3.12}
	1 - \int_{0}^{\tau} \lambda \frac{\left(\lambda q\right)^{n}}{n!} \exp\left(-\lambda q\right) dq = \sum_{i=0}^{n} \frac{\left(\lambda \tau\right)^{i}}{i!} \exp\left(- \lambda \tau\right).
\end{equation} 

In words, if the sum of waiting times for $n + 1$ waiting times is smaller than $\tau$, then it follows that we have observed more than $n$ events occurring at time $\tau$. The equal sign in \eqref{eq.3.11} implies that both states of knowledge have the same truth-value, that is, are equivalent.

Likewise, if the sum of waiting times for $n+1$ events exceeds $\tau$, then it follows that we yet have to observe more than $n$ events occurring at time $\tau$. The equal sign in \eqref{eq.3.12} implies that both states of knowledge have the same truth-value, that is, are equivalent.

The product and and sum rules of Bayesian probability theory are derived by way of consistency constraints, where consistency is operationalized as follows. If there are there two different routes that lead us to the self-same proposition, then these routes should result in the same probability assignment. This then explains the equivalencies \eqref{eq.3.11} and \eqref{eq.3.12}, consistency demands it, \cite{Jaynes03}.    
 
The corollary of the equivalence relation \eqref{eq.3.12} is, see \eqref{eq.2.2},
\begin{align}
	\label{eq.3.13}
	\frac{\left(\lambda \tau\right)^{n}}{n!} \exp\left(- \lambda \tau\right) &= \sum_{i=0}^{n} \frac{\left(\lambda \tau\right)^{i}}{i!} \exp\left(- \lambda \tau\right) - \sum_{i=0}^{n-1} \frac{\left(\lambda \tau\right)^{i}}{i!} \exp\left(- \lambda \tau\right) \notag \\
	&= \int_{0}^{\tau} \lambda \frac{\left(\lambda q\right)^{n-1}}{\left(n-1\right)!} \exp\left(-\lambda q\right) dq - \int_{0}^{\tau} \lambda \frac{\left(\lambda q\right)^{n}}{n!} \exp\left(-\lambda q\right) dq \notag  \\ 
	&=  \int_{0}^{\tau} \lambda \frac{\left(\lambda q\right)^{n-1}}{\left(n-1\right)!} \exp\left(-\lambda q\right) \left(1 - \frac{\lambda q}{n}\right)dq,
\end{align} 
where $n \geq 1$. It may be easily checked that the corollary equivalence \eqref{eq.3.13} does indeed hold.

Now, we may compute the cumulants of the Poisson distribution of $q$ either by way of the evalution of the moments of the probability distribution \eqref{eq.3.8} or by way of the evalution of the moments of the stochastic \eqref{eq.3.5}. 

The cumulants of a given Exponential waiting time is given as:
\begin{align}
	\label{eq.3.14}
	\mu &= \frac{1}{\lambda},   \notag \\
	\sigma &= \frac{1}{\lambda}, \notag \\
	\gamma &= 2, \\
	\kappa &= 9. \notag 
\end{align}
So, the cumulants of $n+1$ exponential waiting times, \eqref{eq.3.5}, are given as:
\begin{align}
	\label{eq.3.15}
	\mu_{n+1} &= \frac{n+1}{\lambda},   \notag \\
	\sigma_{n+1} &= \frac{\sqrt{n+1}}{\lambda}, \notag \\
	\gamma_{n+1} &= \frac{2}{\sqrt{n+1}}, \\
	\kappa_{n+1} &= \frac{9}{n+1} + 3, \notag 
\end{align}

Subtituting the cumulants \eqref{eq.3.15} in the MaxEnt approximative distribution\footnote{See Section~\ref{MaxEnt}.}, with the integral limits 
\begin{equation}
	\label{eq.3.16}
	a = \begin{cases} 
				-\frac{\mu}{\sigma}, \quad &\text{if } \mu - 6 \sigma < 0 \\
				\mu - 6 \sigma < 0,  \quad &\text{else.}
			\end{cases}, \qquad  b = \mu + 6 \sigma,
\end{equation}
for the integral function \eqref{eq.2.14}, and going through the motions, we obtain for, say, $\lambda = 1$ the following MaxEnt distribution:
\begin{figure}[h]
	\centering
		\includegraphics[width=0.70\textwidth]{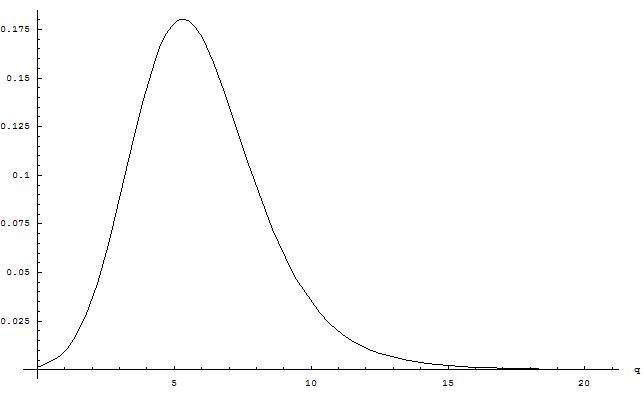}
	\caption{MaxEnt distribution of sum of $n+1$ Exponential waiting times}
	\label{fig:plot1}
\end{figure}

\noindent We can compare Figure~\ref{fig:plot1} with the actual distribution \eqref{eq.3.8}, for $\lambda = 1$:

\begin{figure}[h]
	\centering
		\includegraphics[width=0.70\textwidth]{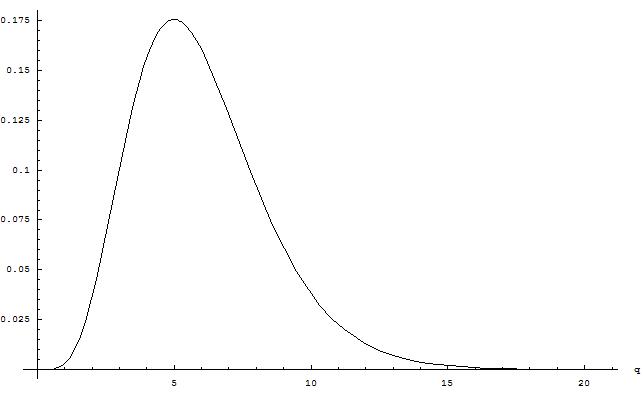}
	\caption{Analytical distrubution of sum of $n+1$ Exponential waiting times}
	\label{fig:plot2}
\end{figure}

By way of the equivalency \eqref{eq.3.13}, the road is now opened to evaluate the first cumulants of \eqref{eq.3.4}. This will allow us construct the MaxEnt approximation, as discussed in Section~\ref{MaxEnt}, of the Beta-Like distribution of a Poisson-Like process, where the event generating mechanism follows a Weibull instead of an Exponential distribution.

\section{Making Use of the Equivalence Relation}
Let $q$ be the sum of $n + 1$ waiting times, which are generated by independent Weibull processes:
\begin{equation}
	\label{eq.3.17}
	q = t_{1} + \cdots + t_{n+1}, \qquad t_{i} \thicksim Weibull\left(k, \lambda\right),
\end{equation}
where the Weibull distribution is given as
\begin{equation}
	\label{eq.3.18}
	p\!\left(\left.t\right|k, \lambda\right) = k \lambda \left(k \lambda\right)^{k-1} \exp\left[\left(- t \lambda\right)^{k}\right].
\end{equation}
The cumulants of a given Weibull waiting time are given as, \eqref{eq.3.18}:
\begin{align}
	\label{eq.3.19}
	\mu &= \int_{0}^{\infty} t \: p\!\left(\left.t\right|k, \lambda\right) dt ,   \notag \\
	\sigma &= \sqrt{\int_{0}^{\infty} \left(t - \mu\right)^{2}  p\!\left(\left.t\right|k, \lambda\right) dt}, \notag \\
	\gamma &= \frac{1}{\sigma^{3}}\int_{0}^{\infty} \left(t - \mu\right)^{3} p\!\left(\left.t\right|k, \lambda\right) dt, \\
\kappa &= \frac{1}{\sigma^{4}}\int_{0}^{\infty} \left(t - \mu\right)^{4} p\!\left(\left.t\right|k, \lambda\right) dt, \notag 
\end{align}
which evaluates to
\begin{align}
	\label{eq.3.20}
	\mu &= \frac{\Gamma\!\left(\frac{1}{k}\right)}{k \lambda},   \notag \\
	\sigma &= \frac{\sqrt{\Gamma\!\left(\frac{k+2}{k}\right) - \Gamma\!\left(\frac{k+1}{k}\right)^{2}}}{\lambda}, \notag \\
	\gamma &= \frac{k^{3} \: \Gamma\!\left(\frac{k+3}{k}\right) - 6 \: k \: \Gamma\!\left(\frac{2}{k}\right) \Gamma\!\left(\frac{1}{k}\right)  + 2 \: \Gamma\!\left(\frac{1}{k}\right)^{3}}{k^{3} \left[\Gamma\!\left[\frac{k+2}{k}\right) - \Gamma\!\left(\frac{k+1}{k}\right)^{2}\right]^{\frac{3}{2}}}, \\
\kappa &= \frac{k^{4} \: \Gamma\!\left(\frac{k+4}{k}\right) -12 \: k^{2} \: \Gamma\!\left(\frac{3}{k}\right) \Gamma\!\left(\frac{1}{k}\right)  + 12 \: k \: \Gamma\!\left(\frac{2}{k}\right) \Gamma\!\left(\frac{1}{k}\right)^{2}   - 3 \: \Gamma\!\left(\frac{1}{k}\right)^{4}}{k^{4} \left[\Gamma\!\left[\frac{k+2}{k}\right) - \Gamma\!\left(\frac{k+1}{k}\right)^{2}\right]^{2}}. \notag 
\end{align}
So, the cumulants of $n+1$ Weibull waiting times, \eqref{eq.3.17}, are given as, \eqref{eq.3.20}:
\begin{align}
	\label{eq.3.21}
	\mu_{n+1} &= \left(n+1\right) \mu,   \notag \\
	\sigma_{n+1} &= \sqrt{n+1} \sigma, \notag \\
	\gamma_{n+1} &= \frac{\gamma}{\sqrt{n+1}}, \\
	\kappa_{n+1} &= \frac{\kappa}{n+1} + 3, \notag 
\end{align}
Likewise, the cumulants of $n$ Weibull waiting times, \eqref{eq.3.17}, are given as, \eqref{eq.3.20}:
\begin{align}
	\label{eq.3.22}
	\mu_{n} &= n \: \mu,   \notag \\
	\sigma_{n} &= \sqrt{n} \sigma, \notag \\
	\gamma_{n} &= \frac{\gamma}{\sqrt{n}}, \\
	\kappa_{n} &= \frac{\kappa}{n} + 3, \notag 
\end{align}

The cumulants \eqref{eq.3.21} and \eqref{eq.3.22} may be used to construct the MaxEnt approximations of, respectively, the distributions of $n+1$ and $n$ Weibull waiting times. These MaxEnt distributions may then substituted into \eqref{eq.3.13}, in order to get an approximation of the intractable Poisson-Like probability model \eqref{eq.3.4}. 

In order to construct the Beta-Like distribution of the Poisson-Like probability model \eqref{eq.3.4}, the MaxEnt distributions that take as their inputs the cumulants in \eqref{eq.3.21} and \eqref{eq.3.22}, which are functions of the unknown parameters $k$ and $\lambda$, \eqref{eq.3.20}, have to be weighted by the Weibull posterior \eqref{eq.1.35}, and summated.

For example, if partition the 6-sigma of $\left(k,\lambda\right)$-parameter space in a $n$-by-$n$ grid, then we substitute the center coordinates of the $\left(k,\lambda\right)$ squares in the cumulants \eqref{eq.3.21} and \eqref{eq.3.22}, construct and weigh the resulting $n\times n = n^{2}$ MaxEnt distributions with probability volumes of the corresponding squares, we then summate these weighed MaxEnt distributions, which will leave us with an approximation of the highly intractable Beta-Like distribution of the Poisson-Like probability model \eqref{eq.3.4}.

Likewise, if we wish to find the Beta-Like distribution of the Poisson-Like equivalence of \eqref{eq.2.20}, then by way of \eqref{eq.3.8}, \eqref{eq.3.11}, and \eqref{eq.3.21}, we may construct, with the above described procedure, of weiging Maxent distributions, its approximative distribution.

\chapter{Bayesian Model Selection}
In Bayesian statistic there are four entitities of interest: the prior, the likelihood, the posterior, and the evidence. Now, anyone somewhat familiar with Bayesian statistics probably knows about the first three of these entitites, since these are needed for Bayesian parameter estimation. However, the fourth entity, the evidence, essential for Bayesian model selection, is less well known.

This is unfortunate. Because, even if the posterior represents the optimal parameter estimation procedure, if the model employed is inappropriate, then the optimality of the parameter estimation procedure will not make the underlying model less inappropriate. And we quote Skilling:
\begin{quotation}
I know of no other discipline in which half of the principal equation is so widely ignored, and it should not be ignored here either. I could (and often do) argue that the evidence 
\[
	Z = p\!\left(D\right) = \int p\!\left(\left.D\right|\lambda\right) d\lambda 
\]	
is even more important than the posterior
\[
p\!\left(\left.D\right|\lambda\right) = \frac{p\!\left(\lambda, D\right)}{p\!\left(D\right)} = \frac{p\!\left(\lambda, D\right)}{Z} 
\]
on the frounds that algebraically it has to be evaluated first, and logically there's no need to proceed to the posterior if the evidence is unacceptably weaker than that from some other candidate. So it's the posterior that is subordinate to the evidence and definitely \emph{not} the other way around. I myself think of ``Bayesian inference'' as the generation of the evidence, with the posterior following if needed. Evidence is primary.  	
\end{quotation}
 
Now, the reason that we have gotten as far as we have, algebraically speaking, without introducing the concept of the evidence, is because we have made use of the fact that the prior time the likelihood is proportional to the posterior: 
\begin{equation}    
 	\label{eq.4.1}
 	\pi\!\left(\lambda\right) L\!\left(\lambda\right) \propto p\!\left(\left.\lambda\right|D\right),
\end{equation}
where $\pi\!\left(\lambda\right)$ is proportional to the prior $p\!\left(\lambda\right)$ and $L\!\left(\lambda\right)$ is proportional to $p\!\left(\left.D\right|\lambda\right)$, the probability of the data given the parameter $\lambda$.

Seeing that any probability distribution should integrate to one, we may properly normalize \eqref{eq.4.1} by way of the indentity:
\begin{equation}    
 	\label{eq.4.2}
 	p\!\left(\left.\lambda\right|D\right) = \frac{\pi\!\left(\lambda\right) L\!\left(\lambda\right)}{\int  \pi\!\left(\lambda\right) L\!\left(\lambda\right) d\lambda} = \frac{\pi\!\left(\lambda\right) L\!\left(\lambda\right)}{C}.
\end{equation}

Note that the normalizing constant
\begin{equation}    
 	\label{eq.4.3}
 	C = \int  \pi\!\left(\lambda\right) L\!\left(\lambda\right) d\lambda,
\end{equation}
is only equal to the evidence 
\begin{equation}    
 	\label{eq.4.4}
 	Z = \int  p\!\left(\lambda\right) p\!\left(\left.D\right|\lambda\right) d\lambda,
\end{equation}
if the $\pi$ and $L$ are both properly normalized; where the former is normalized over the unknown parameter $\lambda$ and the latter over the data $D$.

Note that until now, by way of the use of proportionality sign, we have used the Bayesian short hand \eqref{eq.4.1} to present our posteriors. In what follows, We will compute, for demonstrative purposes, the evidences for the models in which the generating failure mechanisms are Exponential and Weibull, respectively. But first we will give a simple outline of the procudure of Bayesian model selection.

\section{Bayesian Model Selection}
Let $p\!\left(\left.\lambda\right|I\right)$ be the prior of some parameter $\lambda$, conditional on the prior background information $I$. Let $p\!\left(\left.D\right|\lambda,M\right)$ be the probability of the data $D$, conditional on a given parameter $\lambda$ and the particular likelihood model $M$ which was invoked. Let $p\!\left(\left.\lambda\right|D, M, I\right)$ be the posterior distribution of $\lambda$, conditional on the data $D$, the likelihood model $M$, and the prior background information $I$. 

We then have, by way of the product rule, or, equivalently, Bayes' theorem, that, \cite{Jaynes03}: 
\begin{equation}    
 	\label{eq.4.5}
 	p\!\left(\left.\lambda\right|D, M, I\right) = \frac{p\!\left(\left.\lambda\right|I\right) p\!\left(\left.D\right|\lambda,M\right)}{p\!\left(\left.D\right|M, I\right)},
\end{equation}
where
\begin{equation}    
 	\label{eq.4.6}
 	p\!\left(\left.D\right|M, I\right) = \int p\!\left(\left.\lambda\right|I\right) p\!\left(\left.D\right|\lambda,M\right) d\lambda,
\end{equation}
is the marginalized likelihood of the model $M$ and the background information $I$, also known as the evidence of $M$ and $I$.

Note that the evidence judges \eqref{eq.4.6} judges both the likelihood model, $M$, as well as the prior model, $I$, that went into the construction of the posterior. Now, this could be seen as a weakness of Bayesian model selection\footnote{As was once suggested to the first author, during a colloquiem on Bayesian model selection.}, since in general all the ingenuity goes into the construction of a sophisticated likelihood model. So, why bother with some uninformative prior, when we only want to compare different likelihood models?

There are two reasons why it is a good thing that Bayesian model selection takes into account both the prior and the likelihood, and does not neglect the former. 

Firstly, there are instances, for example in image reconstruction \cite{Skilling84}, where all the artfulness goes into the construction of the prior, and there we have that it is the likelihood which is trivial. So, it is precisely because the Bayesian probability theory puts the prior and likelihood models automatically on an equal footing, that Bayesian model selection can differentiate between the different prior models of image reconstruction inference problems, without breaking down.

Secondly, by judging the prior the evidence automatically guards us against the danger of over-parametrization, that is, choosing such a complex likelihood model, in terms of the number of parameters employed, that we fit the noise in the data as structural part of the data.

Say we have $m$ different likelihood models, $M_{1}, \ldots, M_{m}$, to choose from and one class of, say, uninformative prior background models, $I$. Then we may compute $m$ different evidence values $p\!\left(\left.D\right|M_{j}, I\right)$, for $j = 1, \ldots, m$. 

Let $p\!\left(M_{j}\right)$ be the prior probability distribution of the likelihood models $M_{j}$, and let $p\!\left(\left.M_{j}\right| D, I\right)$ be the posterior probability distribution of these models, conditional on the data and the general prior background information $I$. Then we have that
\begin{equation}    
 	\label{eq.4.7}
 	p\!\left(\left.M_{j}\right|D, I\right) = \frac{p\!\left(M_{j}\right) p\!\left(\left.D\right|M_{j}, I\right)}{\sum_{i} p\!\left(M_{i}\right) p\!\left(\left.D\right|M_{i}, I\right)}
\end{equation}
for $j = 1, \ldots, m$.

Note that if we have that $p\!\left(M_{j}\right) = p\!\left(M_{k}\right)$, for $j \neq k$, then we have that  \eqref{eq.4.7} reduces to
\begin{equation}    
 	\label{eq.4.8}
 	p\!\left(\left.M_{j}\right|D, I\right) = \frac{p\!\left(\left.D\right|M_{j}, I\right)}{\sum_{i} p\!\left(\left.D\right|M_{i}, I\right)}
\end{equation}
Stated differently, if we assign equal prior probabilities to our different likelihoods models, then the posterior probabilities of these likelihood models reduce to their normalized evidence values. This, then, is why the likelihood models may be ranked by their respective evidence values, \cite{MacKay03}.

\section{Computing Evidence Values}
For the Exponential model, which we designate $M_{1}$, we have that the likelihood model is given as, \eqref{eq.1.18},
\begin{align}
	\label{eq.4.9}
	p\!\left(\left.D\right|\lambda, M_{1}\right) &= \prod_{i=1}^{r} \lambda \exp\left(-\lambda x_{i}\right) dx_{i} \prod_{j=1}^{n-r} \exp\left(-\lambda y_{j}\right) \notag \\
	&=  \lambda^{r} \exp\left[-\lambda \left(\sum_{i=1}^{r} x_{i} + \sum_{j=1}^{n-r} y_{j} \right)\right] \prod_{i=1}^{r} dx_{i}.
\end{align} 
As a prior we take the properly normalized uninformative prior:
\begin{equation}
	\label{eq.4.10}
	p\!\left(\left.\lambda\right|I\right) = \frac{C_{\lambda}}{\lambda},
\end{equation} 
where $C_{\lambda}$ is the normalizing constant
\begin{equation}
	\label{eq.4.11}
	C_{\lambda}^{-1} = \int_{a_{\lambda}}^{b_{\lambda}} \frac{d \lambda}{\lambda} = \log b_{\lambda} - \log a_{\lambda}. 
\end{equation} 
where $a_{\lambda}$ and $b_{\lambda}$ define the prior range of possible values of $\lambda$. 

Cogent prior information regarding $a_{\lambda}$ is that value of $\lambda$ for which, for some given time interval $tau$, the expection value $\lambda \tau$ would become so small that too few failures would be witnessed in said time period. Cogent prior information regarding $b_{\lambda}$ is that value of $\lambda$ for which, for some given time interval $tau$, the expection value $\lambda \tau$ would become so large that too many failures would be witnessed in said time period.

Multiplying the properly normalized likelihood \eqref{eq.4.9} with the the properly normalized prior \eqref{eq.4.10}, we obtain the the properly normalized bivariate distribution of both the parameter and the data:
\begin{equation}
	\label{eq.4.12}
	p\!\left(\left.\lambda, D\right|M_{1}, I\right) =  C_{\lambda}\lambda^{r-1} \exp\left[-\lambda \left(\sum_{i=1}^{r} x_{i} + \sum_{j=1}^{n-r} y_{j} \right)\right] \prod_{i = 1}^{r} dx_{i},
\end{equation} 
which, being properly normalized, will allow us to evaluate the evidence of $M_{1}$.

By way of \eqref{eq.4.6} and \eqref{eq.4.12}, we then evaluate the evidence for the Exponential model as 
\begin{align}
	\label{eq.4.13}
		p\!\left(\left.D\right|M_{1}, I\right) &= \int_{a_{\lambda}}^{b_{\lambda}} p\!\left(\left.\lambda,D\right|M_{1}, I\right) d\lambda \notag \\
		&=  C_{\lambda} \prod_{i} dx_{i} \int_{a_{\lambda}}^{b_{\lambda}} \lambda^{r-1} \exp\left[-\lambda \left(\sum_{i} x_{i} + \sum_{j} y_{j} \right)\right] d\lambda \\
		&\approx C_{\lambda}  \frac{\left(r-1\right)!}{\left(\sum_{i} x_{i} + \sum_{j} y_{j}\right)^{r}} \prod_{i} dx_{i}. 
\end{align} 

For the Weibull model, which we designate $M_{2}$, we have that the likelihood model is given as, \eqref{eq.1.31},
\begin{align}
	\label{eq.4.14}
	p\!\left(\left.D\right|k, \lambda, M_{2}\right) &= \prod_{i=1}^{r} k \lambda \left(\lambda x_{i}\right)^{k-1} \exp\left[\left(-\lambda x_{i}\right)^{k}\right] d x_{i} \prod_{j=1}^{n-r} \exp\left[\left(-\lambda y_{j}\right)^{k}\right] \notag \\
	&= \lambda^{r k} k^{r} \exp\left[- \lambda^{k} \left(\sum_{i=1}^{r}  x_{i}^{k} +\sum_{j=1}^{n-r}  y_{j}^{k}\right)\right] \prod_{i=1}^{r} x_{i}^{k-1} dx_{i}.
\end{align}   

As a prior we take the properly normalized uninformative prior: 
\begin{align}
	\label{eq.4.15}
	p\!\left(\left.k, \lambda\right|I\right) &= p\!\left(\left.k\right|I\right) p\!\left(\left.\lambda\right|I\right) \notag \\
																					 &= \frac{C_{k}}{k} \frac{C_{\lambda}}{\lambda}
\end{align} 
where $C_{\lambda}$ is as in \eqref{eq.4.11} and $C_{k}$ is the normalizing constant
\begin{equation}
	\label{eq.4.16}
	C_{k}^{-1} = \int_{a_{k}}^{b_{k}} \frac{d k}{k} = \log b_{k} - \log a_{k}. 
\end{equation} 
where $a_{k}$ and $b_{k}$ define the prior range of possible values of $k$. 

Multiplying the properly normalized likelihood \eqref{eq.4.14} with the the properly normalized prior \eqref{eq.4.15}, we obtain the the properly normalized bivariate distribution of both the parameters and the data:
\begin{equation}
	\label{eq.4.17}
	p\!\left(\left.k, \lambda, D\right|M_{2}, I\right) =  C_{\lambda} C_{k} \lambda^{r k - 1} k^{r - 1} \exp\left[- \lambda^{k} \left(\sum_{i=1}^{r}  x_{i}^{k} +\sum_{j=1}^{n-r}  y_{j}^{k}\right)\right] \prod_{i=1}^{r} x_{i}^{k-1} dx_{i},
\end{equation} 
which, being properly normalized, will allow us to evaluate the evidence of $M_{2}$.

By way of \eqref{eq.4.6} and \eqref{eq.4.17}, we then evaluate the evidence for the Weibull model as 
\begin{align}
	\label{eq.4.18}
		p\!\left(\left.D\right|M_{2}, I\right) &= \int_{a_{k}}^{b_{k}} \int_{a_{\lambda}}^{b_{\lambda}}  p\!\left(\left.k, \lambda,D\right|M_{2}, I\right) d\lambda \: dk  \notag \\
		&=  C_{\lambda} C_{k} \prod_{i} dx_{i} \int_{a_{k}}^{b_{k}} \int_{a_{\lambda}}^{b_{\lambda}}  \lambda^{r k - 1} k^{r - 1} \exp\left[- \lambda^{k} \left(\sum_{i}  x_{i}^{k} +\sum_{j}  y_{j}^{k}\right)\right] \prod_{i} x_{i}^{k-1} \: d\lambda \: dk \notag \\
		&\approx C_{\lambda} C_{k}  \: \left(r-1\right)! \: \prod_{i} dx_{i}  \: \int_{a_{k}}^{b_{k}} \frac{k^{r-2} \prod_{i} x_{i}^{k-1}}{\left(\sum_{i}  x_{i}^{k} +\sum_{j}  y_{j}^{k}\right)^{r}}  \: dk,
\end{align} 
where the integral over unknown shape parameter $k$ must be evaluated numerically.

Say, we do not have any prior preference for either model $M_{1}$ or $M_{2}$. Then, letting the data speak for itself, we assign equal prior probabilities to both likelihood models. We then, by way of \eqref{eq.4.8}, \eqref{eq.4.13}, and \eqref{eq.4.18}, may compute the posterior probability of $M_{1}$:
\begin{align}
	\label{eq.4.19}
		p\!\left(\left.M_{1}\right|D, I\right) &= \frac{p\!\left(\left.D\right|M_{1}, I\right)}{p\!\left(\left.D\right|M_{1}, I\right) + p\!\left(\left.D\right|M_{2}, I\right)} \notag \\
		&\approx \frac{\left[C_{\lambda} \left(r-1\right)! \prod_{i} dx_{i}\right] \frac{1}{\left(\sum_{i} x_{i} + \sum_{j} y_{j}\right)^{r}}}{\left[C_{\lambda} \left(r-1\right)! \prod_{i} dx_{i}\right]  \frac{1}{\left(\sum_{i} x_{i} + \sum_{j} y_{j}\right)^{r}} + \left[C_{\lambda} \left(r-1\right)! \prod_{i} dx_{i} \right] C_{k} \int \frac{k^{r-2} \prod_{i} x_{i}^{k-1}}{\left(\sum_{i}  x_{i}^{k} +\sum_{j}  y_{j}^{k}\right)^{r}} \: dk} \notag \\
		&= \frac{\left(\sum_{i} x_{i} + \sum_{j} y_{j}\right)^{-r}}{\left(\sum_{i} x_{i} + \sum_{j} y_{j}\right)^{-r} + C_{k} \int \frac{k^{r-2} \prod_{i} x_{i}^{k-1}}{\left(\sum_{i}  x_{i}^{k} +\sum_{j}  y_{j}^{k}\right)^{r}} \: dk}, 
\end{align} 
where we have cancelled out all the terms shared by both evidence values \eqref{eq.4.13} and \eqref{eq.4.18}. 

Furthermore, seeing that, the normalizing constant $C_{\lambda}$, \eqref{eq.4.11}, cancels out, we may let \eqref{eq.4.10} be an improper prior and let the constants of integration go to $a_{\lambda} \rightarrow 0$ and $b_{\lambda} \rightarrow \infty$. This allows us to replace the `approximately-equal-to' signs in \eqref{eq.4.13} and \eqref{eq.4.18} with an equality signs, which propagates through in \eqref{eq.4.19}.

By way of \eqref{eq.4.8}, \eqref{eq.4.13}, and \eqref{eq.4.18}, we may also compute the posterior probability of $M_{2}$:
\begin{align}
	\label{eq.4.20}
		p\!\left(\left.M_{2}\right|D, I\right) &= \frac{p\!\left(\left.D\right|M_{2}, I\right)}{p\!\left(\left.D\right|M_{1}, I\right) + p\!\left(\left.D\right|M_{2}, I\right)} \notag \\
		&\approx \frac{\left[C_{\lambda} \left(r-1\right)! \prod_{i} dx_{i} \right] C_{k} \int \frac{k^{r-2} \prod_{i} x_{i}^{k-1}}{\left(\sum_{i}  x_{i}^{k} +\sum_{j}  y_{j}^{k}\right)^{r}} \: dk}{\left[C_{\lambda} \left(r-1\right)! \prod_{i} dx_{i}\right]  \frac{1}{\left(\sum_{i} x_{i} + \sum_{j} y_{j}\right)^{r}} + \left[C_{\lambda} \left(r-1\right)! \prod_{i} dx_{i} \right] C_{k} \int \frac{k^{r-2} \prod_{i} x_{i}^{k-1}}{\left(\sum_{i}  x_{i}^{k} +\sum_{j}  y_{j}^{k}\right)^{r}} \: dk} \notag \\
		&= \frac{C_{k} \int \frac{k^{r-2} \prod_{i} x_{i}^{k-1}}{\left(\sum_{i}  x_{i}^{k} +\sum_{j}  y_{j}^{k}\right)^{r}} \: dk}{\left(\sum_{i} x_{i} + \sum_{j} y_{j}\right)^{-r} + C_{k} \int \frac{k^{r-2} \prod_{i} x_{i}^{k-1}}{\left(\sum_{i}  x_{i}^{k} +\sum_{j}  y_{j}^{k}\right)^{r}} \: dk}, 
\end{align} 
where we have again cancelled out all the terms shared by both evidence values \eqref{eq.4.13} and \eqref{eq.4.18}. 

\section{A Word of Caution}
Note that for $\lambda$ we, eventually, used an improper uninformative prior. We did so when we let constants of integration in \eqref{eq.4.11} go to $a_{\lambda} \rightarrow 0$ and $b_{\lambda} \rightarrow \infty$; thus, giving us a normalizing constant of 
\begin{equation}
	\label{eq.4.21}
	C_{\lambda}^{-1} = \int_{0}^{\infty} \frac{d\lambda}{\lambda} = \infty,
\end{equation}
or, equivalently,
\begin{equation}
	\label{eq.4.22}
	C_{\lambda} = \frac{1}{\infty} = 0.
\end{equation}

The only reason we had the freedom to do so was because the constant $C_{\lambda}$ cancelled out in \eqref{eq.4.19} and \eqref{eq.4.20}. Thus, removing the inverse infinities, which resulted from the improper, that is, diverging, prior of $\lambda$. Now, had we done the same for constants of integration in \eqref{eq.4.16}, then we would have obtained an inverse infinity $C_{k} = 0$, which would have not cancelled out in \eqref{eq.4.19} and \eqref{eq.4.20}; thus giving us posterior model probabilities:
\begin{equation}
	\label{eq.4.23}
		p\!\left(\left.M_{1}\right|D, I\right) = \frac{\left(\sum_{i} x_{i} + \sum_{j} y_{j}\right)^{-r}}{\left(\sum_{i} x_{i} + \sum_{j} y_{j}\right)^{-r} }  = 1, 
\end{equation}    
and
\begin{equation}
	\label{eq.4.24}
		p\!\left(\left.M_{2}\right|D, I\right) = \frac{0}{\left(\sum_{i} x_{i} + \sum_{j} y_{j}\right)^{-r}} = 0.
\end{equation} 

We see in \eqref{eq.4.23} and \eqref{eq.4.24} how Bayesian model selection may punish us for non-parsimoneous priors when the normalizing constants of these priors do not cancel out in the posterior of the competing likelihood models. To the uninitiated this may seem as a bother. But we Bayesians would not have it any other way. Because it is this penalizing mechanism of Bayesian model selection, which is just a straight forward of the product and sum rules, which automatically protects us from the dangers of over-fitting. 

For example, these authors had to choose among regression models\footnote{Those regression models being C-splines models, \cite{vanErp11}.} having four up to a thousand possible regression coefficients to model noisy data. By deriving a parsimoneous prior for the regression coefficients,\cite{vanErp14}, we were able to let the data do the talking. We found that the Bayesian probability theory picked the likelihood model having only sixty-four parameters. Those models having more parameters, though having a better likelihood fit, because of the greater number of parameters, were penelized for their prior probability volume and, as consequence, noise was minimally fitted as part of the structure.

The take-home message from all this is the following: In the computing of the evidences, \eqref{eq.4.6},
\begin{enumerate}
	\item improper priors should only be used if their normalizing constants will cancel out in \eqref{eq.4.7}, and
	\item priors whose normalizing constants do not cancel out should be as parsimoneous as possible.
\end{enumerate}


\begin{thebibliography}{9}

\bibitem {vanErp11}
Erp~van H.R.N., Linger R.O., and Gelder~van P.H.A.J.M.:  \textit{Constructing Cartesian Splines}. The Open Numerical Methods Journal, 3, 26-30, (2011). But we recommend to search for the unmutilated arXiv version of this article: arXiv:1409.5955 [math.NA], (2014).

\bibitem {vanErp13b}
Erp~van H.R.N., Linger R.O., and Gelder~van P.H.A.J.M.: \textit{Deriving Proper Uniform Priors for Regression Coefficients, Part II}, arXiv:1308.1114 [stat.ME], (2013).

\bibitem {vanErp14}
Erp~van H.R.N., Linger R.O., and Gelder~van P.H.A.J.M.: \textit{Fact Sheet Reseach on Bayesian Decision Theory}, arXiv, (2014).

\bibitem {Hall92}
Hall P.: \textit{The Bootstrap and Edgeworth Expansion}, Springer-Verlag (1992).

\bibitem {Jaynes76}
Jaynes E.T.: \textit{Confidence Intervals vs Bayesian Intervals; Reply to Kempthorne's Comments}, W.L. Harper and C.A. Hooker, eds. Foundations of Probability Theory, Statistical Inference, and Statistical Theories of Science, Reidel Publishing Co., Dordrecht, Holland, (1976).

\bibitem {Jaynes03}
Jaynes E.T.: \textit{Probability Theory; the Logic of Science}. Cambridge University Press, (2003).

\bibitem {MacKay03}
MacKay D.J.C.: \textit{Information Theory, Inference, and Learning Algorithms}, Cambridge University Press, Cambridge, (2003).

\bibitem {Skilling84}
Skilling J.: \textit{Fundamentals of MaxEnt in Data Analysis}, Buck B.B. and Macauly V.A, eds. Maximum Entropy in Action, Clarendon Press, Oxford, (1991).

\bibitem {Rockinger01}
Rockinger and Jondeau: \textit{Entropy Densities With an Application to Autoregressive Conditional Skewness and Kurtosis}, Journal of Econometrics 106, 119-142, (2002).

\bibitem {Zellner72}
Zellner A.: \textit{An Introduction to Bayesian Inference in Econometrics}, John Wiley and Sons, Inc., (1971).

\end{thebibliography}
\end{document}